\documentstyle[12pt,a4,epsf]{article}
\newcommand{\rcite}[1]{{\cite{#1}}}
\newcommand{\rref}[1]{{(\ref{#1})}}
\newcommand{\tref}[1]{{\ref{#1}}}
\newcommand{\rlabel}[1]{{\label{#1}}}
\newcommand{\rbibitem}[1]{\bibitem{#1}}
\newcommand{\be}{\begin{equation}}
\newcommand{\ee}{\end{equation}}
\newcommand{\ba}{\begin{eqnarray}}
\newcommand{\ea}{\end{eqnarray}}
\newcommand{\dis}{\displaystyle}

\newcommand{\Pids}{\Pi_{\Delta S =2}}
\newcommand{\mathrm}[1]{{\rm #1}}

\begin{document}
\begin{titlepage}
\begin{flushright}
{NORDITA-94/44 N,P\\revised}
\end{flushright}
\vspace{2cm}
\begin{center}
{\large\bf $B_K$ and Explicit Chiral Symmetry Breaking}\\
\vfill
{\bf Johan Bijnens$^a$ and Joaquim Prades$^{a,b}$}\\[0.5cm]
${}^a$ NORDITA, Blegdamsvej 17,\\
DK-2100 Copenhagen \O, Denmark\\[0.5cm]
$^b$ Niels Bohr Institute, Blegdamsvej 17,\\
DK-2100 Copenhagen \O, Denmark
\end{center}
\vfill
\begin{abstract}
The $B_K$ parameter is discussed in the general context of calculating
beyond the factorization approximation for hadronic matrix elements.
A variant of the $1/N_c$ method of Bardeen et al. is used. We present
calculations within a low energy approach and a calculation within the
Nambu--Jona-Lasinio model. Matching with the QCD behaviour and dependence
on non-zero current quark masses is studied.
\end{abstract}
\vfill
September 1994
\end{titlepage}
\section{Introduction}
The problem of calculating matrix elements of weak decay operators is a
rather old one. In this letter we will restrict ourselves to the
matrix element between two on-shell kaon states of the following
$\Delta S=2$ four-quark operator
\be
{\cal O}_{\Delta S=2} (x)  \equiv L^{sd}_\mu(x) L_{sd}^\mu (x)
\ee
with $L^{sd}_\mu(x) = \overline s (x) \gamma_\mu
\left(\frac{\dis 1-\gamma_5}{\dis 2}\right) d (x)$ and
summation over colours is understood.
This matrix element is usually parametrized in the form of
the $B_K$-parameter times the vacuum insertion approximation
(VIA) as follows
\be
\rlabel{defbk}
\langle \overline K^0  | {\cal O}_{\Delta S=2} (x)
| K^0  \rangle \equiv \frac{\dis 4}{\dis 3}
B_K (\mu) f^2_K m_K^2
\ee
where $f_K$ denotes the $f_K \to \mu \nu$ coupling ($f_K =
113$ MeV in this normalization) and $m_K$ is the $K^0$ mass.
The $\mu$-scale dependence of $B_K$ reflects the fact that
the four-quark operator ${\cal O}_{\Delta S =2}$ has an
anomalous dimension and its matrix element runs between
the scale where the operator has a hadronic realization
and the scale where it matches the quark-gluon
realization of QCD. This anomalous dimension
is known and using the renomalization group
leads to the definition of the scale invariant quantity
\ba
\hat B_K & =& B_K (\mu) \alpha_S (\mu)^{a_+}
\ea
with
\be
a_+ = \frac{\dis 3}{\dis -2 \beta^{(1)}}
\left(1-\frac{\dis 1}{\dis N_c}\right)
\ee
at one-loop. Here $\beta^{(1)}$ is the QCD beta function first
 coefficient. For three active quark flavours and $N_c =3$ we have
$a_+ =-2/9$.

The vacuum insertion approximation
was historically the first way this particular matrix element was evaluated
\rcite{GAILLARD}.
Here by definition we have $B_K (\mu) =1$ at any scale
and we can only obtain an order of magnitude estimate.
Next this matrix element was related to the $\Delta I=3/2$ part of
$K\to\pi\pi$ by Donoghue et al.\rcite{Donoghue1}
using SU(3) symmetry and PCAC. This leads to a value
$\hat B_K \approx 0.37$. It was then found that this relation has rather large
corrections\rcite{BSW} due to SU(3) breaking. Then three new
analytical approaches appeared, the hadronic duality approach
\rcite{AP1}, QCD sum rules
using three-point functions\rcite{QCD1} and the $1/N_c$ expansion
\rcite{BBG1}. The lattice QCD method
also started producing preliminary results around this time.
A review of the situation several years ago can be found in the
proceedings of the Ringberg workshop devoted to this subject\rcite{Buras1}.
All these approaches have in common that they try to get a numerical value
for the $B_K$ parameter and study its dependence on the renormalization
scale $\mu$.
All of these methods have been updated and refined. The hadronic duality
update can be found in \rcite{prades1}, a QCD sum rule calculation
is in \rcite{QCD2} and the $1/N_c$ expansion method has had the vector meson
contribution calculated in a Vector Meson Dominance (VMD)
model\rcite{Gerard}. A review of recent lattice
results can be found in \rcite{lattice}. A full Chiral Perturbation
Theory (CHPT) approach to the problem is unfortunately
not possible. The data on kaon nonleptonic decays do not allow
to determine all relevant parameters at next-to-leading order
(${\cal O} (p^4)$) in
the nonleptonic chiral Lagrangian\rcite{kambor}.
A calculation of these parameters within a QCD inspired model
can be found in \rcite{Bruno} where the determination of the
$B_K$ factor is done to ${\cal O} (p^4)$.

The  leading order
result for $B_K$ in the $1/N_c$ expansion is well known
\be
\rlabel{Nc}
B_K (\mu) = \hat B_K = \frac{\dis 3}{\dis 4} .
\ee
This result is model independent. However, to go further in the
$1/N_c$ expansion requires some model dependent assumptions.
Different low-energy models are then used in variants on
the $1/N_c$ method\rcite{BBG1}. An example is the calculation done
within the QCD-effective action model\rcite{AP2}.
In fact the $\pi^+-\pi^0$ mass difference has been used to
test all of the different $1/N_c$
methods, the original one including vector
mesons \rcite{BBG2}, the QCD-effective action model
\rcite{BR} and the extended Nambu-Jona-Lasinio (ENJL)
model\rcite{BRZ}. The advantage of the latter model
is that it has been shown to give a good picture of the parameters
in the chiral Lagrangian including the SU(3) breaking ones,
see \rcite{BBR} and references therein.
In fact this model accounts for a rather large body of low energy hadronic
data, see \rcite{NJL} for a review.
The reason to resort to a model of this type is that if we also would like
to know possible SU(3) breaking effects, we cannot use a simple vector
meson model because we need to know the SU(3) breaking of couplings of the
resonances to external currents. In practice a large part of these effects
comes from scalar resonances which are experimentally rather poorly
known. We could then use a model to estimate these couplings but it
seems to make more sense then to calculate the relevant quantity
directly within the model. It is also important to know the higher
order couplings beyond order $p^4$ since weak matrix elements
involve an integration over internal momenta.

In this letter we shall give the results for the
next-to-leading corrections to \rref{Nc} for the simple NJL model without
a vector like coupling. This does not include vector mesons so it will
not be the final answer but one can already do a study of the SU(3)
breaking effects since these tend to come mainly from the scalar sector.
Results on the full model and more details regarding the present work will
be presented elsewhere\rcite{BP1}. The full model requires inclusion
of anomalous diagrams and involves about an order of magnitude more terms
than the present work.

We will first discuss the method and then use CHPT
to derive first results. This is essentially the method of \rcite{BBG1}
in a slightly different notation. This will also allow us to test
at low cut-offs the calculation done afterwards. Next we describe the
NJL model and give the lowest order result in $1/N_c$. Then we describe the
main part of the work. The calculation of the non factorizable part of the
$\Delta S=2$ two-point function in the NJL model.
Next we discuss how the Fierz-contribution is reproduced analytically
in this approach and how it affects the final result.
We will give numerical results for the case of
massless quarks, of non-zero
degenerate quark masses and of different quark masses.
In the end we present our main conclusions.

\section{The method and definitions}

We calculate here not directly the $B_K$-factor but the
$\Delta S=2$ two-point function
\ba
\label{twopoint}
G_F\,\Pids(q^2) & \equiv &
i \int d^4 x \, e^{iq\cdot x} \, e^{i \Gamma_{\Delta S =2}}
\langle 0 | T \left( P^{ds}(0)P^{ds}(x)\right)| 0 \rangle
\nonumber \\ &
=& i^2 \int d^4 x \, e^{iq\cdot x}
\langle 0 | T \left( P^{ds}(0)P^{ds}(x) \Gamma_{\Delta S =2}
\right)| 0 \rangle
\ea
in the presence of strong interactions. We shall use the NJL model
for scales
below or around the spontaneous symmetry breaking scale. Here
$G_F$ is the Fermi coupling constant, we use
$P^{ds}(x) = \overline d (x)i\gamma_5 s (x)$, with summation over colour
understood and
\be
\rlabel{operator}
\Gamma_{\Delta S=2} = - G_F\,
\int d^4 y \, {\cal O}_{\Delta S = 2} (y) .
\ee
 The reason to calculate this two-point function rather than
directly the matrix element is that we can now perform the calculation
fully in the Euclidean region so we do not have the problem
of imaginary scalar products. This also allows us in principle to
obtain an estimate of off-shell effects in the matrix elements. This
will be important in later work to assess the uncertainty when trying
to extrapolate from $K\to\pi$ decays to $K\to 2\pi$.
This quantity is also very similar to what is used in
the lattice and QCD sum rule calculations of $B_K$.

The $\Delta S = 2$ operator in \rref{operator} can be
rewritten as
\ba
\rlabel{operator3}
\Gamma_{\Delta S=2} &=& - G_F \,
\int d^4 y L^{sd}_\mu (y) L_{sd}^\mu (y)
\nonumber \\
&=& - G_F \, \int \frac{d^4 r}{(2\pi)^4}
\int d^4 x_1 \int d^4 x_2 \,
e^{-i r \cdot(x_2 - x_1)} L^{sd}_\mu(x_1) L_{sd}^\mu(x_2) .
\ea
This allows us to consider this operator as
being produced at the $M_W$ scale
by the exchange of a heavy $X$ $\Delta S = 2$ boson.
We will work in the Euclidean domain
where all momenta squared are negative.
The integral in the modulus of the momentum $r$ in
\rref{operator3} is then split into two parts,
\be
\int_0^{M_W} d |r| = \int_0^\mu d |r| + \int_\mu^{M_W} d |r| \ .
\ee
In principle one should then evaluate both parts separately as was done
for the $\pi^+-\pi^0$ mass difference
in the above quoted references. Here we will do
the upper part of the integral using the renormalization group.
This results in the integral being of the same form but multiplied
with the Wilson coefficient $C(\mu)$,
\be
\rlabel{operator2}
\Gamma_{\Delta S=2} = - G_F \, C(\mu) \int_0^\mu \frac{d^4 r}{(2\pi)^4}
\int d^4 x_1 \int d^4 x_2 \, e^{-ir\cdot(x_2 - x_1)}
L^{sd}_\mu(x_1) L_{sd}^\mu (x_2)
\ee
A strict analysis in $1/N_c$ would correspond to
set $C(\mu) = 1 + d(\mu) \alpha_S(\mu)$ and to evaluate the second term using
factorization in leading $1/N_c$.
$d(\mu)$ is what the one-gluon exchange diagrams would give with a lower
cut-off $\mu$.
We will however use the full one-loop
Wilson coefficient $C(\mu)=\left(\alpha_S(\mu)/\alpha_S(M_W)\right)^{a_+}$.
In the remainder formulas this factor
will be suppressed for simplicity.
The lower part of the integral is then evaluated using the NJL model.

To illustrate the procedure let us calculate the relevant quantities
in chiral perturbation theory.
This will only make sense for a scale $\mu$ such that
CHPT can be trusted. The lowest order (LO)
contribution comes from the diagram in fig. \tref{fig1}a.
\begin{figure}
\epsfxsize=14cm
\epsfbox{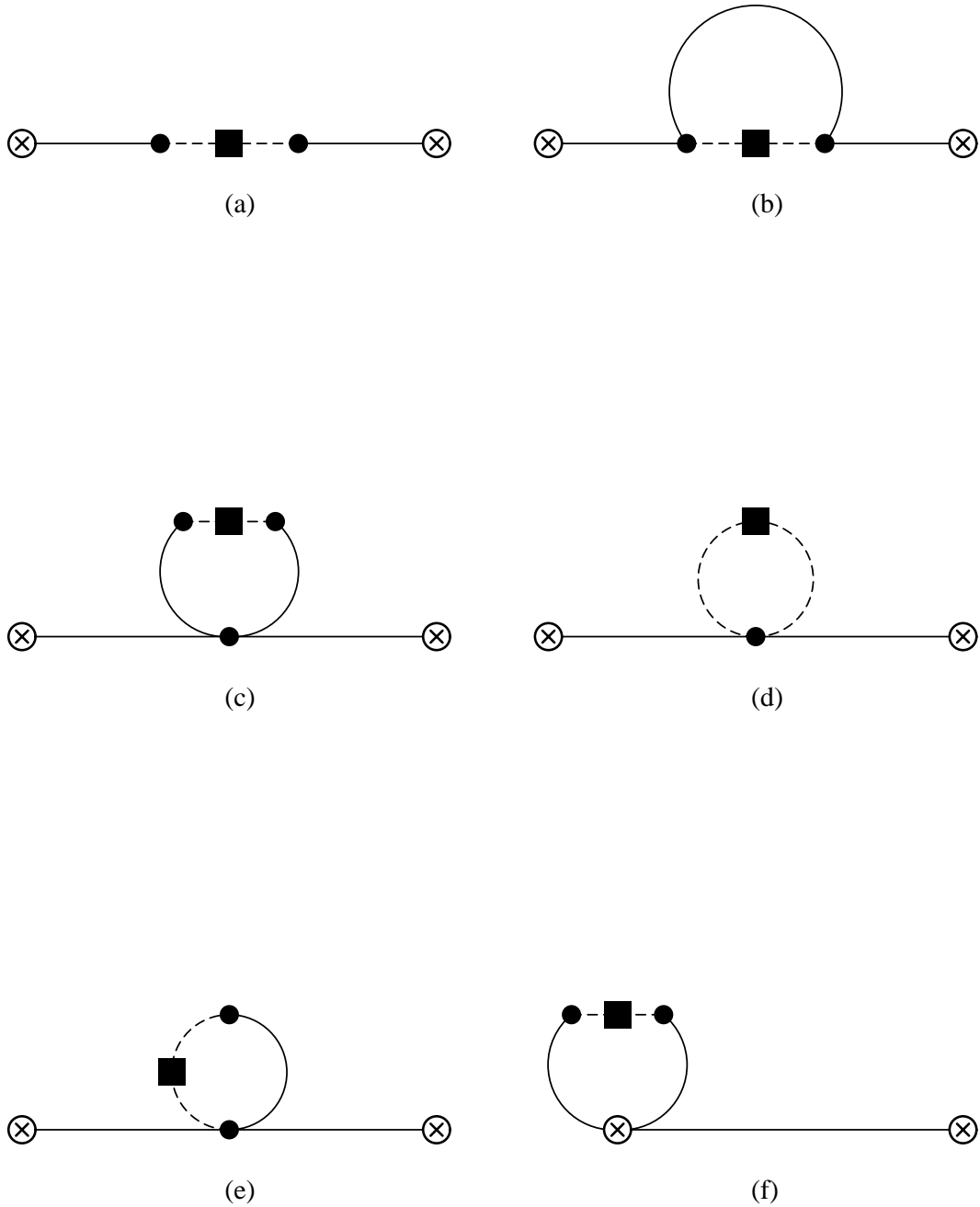}
\caption{Chiral Perturbation Theory contributions to $\Pids(q^2)$.
(a) Lowest order.
(b-f) Higher order non-factorizable. A circled cross
is an insertion of the external pseudoscalar current, $P^{sd}(x)$,
 a dot is a strong
interaction vertex and the square represents the $\Delta S=2$ operator,
$\Gamma_{\Delta S=2}$.
The full lines are meson lines.}
\rlabel{fig1}
\end{figure}
The vertices at this order can be obtained from the standard chiral
Lagrangian of order $p^2$.
The result is
\be
\rlabel{CHPT-LO}
\Pids(q^2)_{CHPT-LO} =
- 2 B_0^2 F_0^4 \frac{q^2}{\left(q^2 - m_K^2\right)^2}\ .
\ee
The parameter $B_0$ is related to the quark condensate in the chiral limit
by $\langle\overline{q}q\rangle = - B_0 F_0^2$ and $F_0$ is the pion decay
constant in the chiral limit, $F_0 \approx 86$ MeV. At this order, the
only nonvanishing contributions are leading in $1/N_c$, therefore
when the result in \rref{CHPT-LO} is properly reduced (see \rcite{BP2})
and compared with eq. \rref{defbk} one gets the
leading order $1/N_c$ result in eq. \rref{Nc}.

At next-to-leading order (NLO) in the chiral expansion
more diagrams have to be taken into account.
The ones that are non-factorizable (NF), i.e.,
the ones that are not reducible when the $X$ boson
propagator gets cut, are depicted
in fig. \tref{fig1}b-f (plus the last one symmetrized).
Here the vector-vector part of the interaction
contributes in diagrams b and d while the axial-axial part contributes in
c,d,e and f. Diagram d was not considered in the usual $1/N_c$ approach.
We will discuss its significance later.
We have to identify the cut-off $\mu$ uniquely
from diagram to diagram. We would also like to get
the model independent leading $1/N_c$ result
independent of the value of the scale $\mu$. We do this by
routing the external momentum ($q$) and the loop momentum ($r$)
through the $X$ boson propagator ($p_X = q + r$ )
and then cutting on the loop momentum in the Euclidean
by $r^2_E\le \mu^2$. Now, in order to
compare with the QCD realization in terms of quarks and gluons
identifying $\mu$ as the renormalization scale,
this then requires $q^2/\mu^2$ and  $\Lambda^2_{QCD}/\mu^2$ to be  small.
In the chiral limit (massless quarks) all integrals can
be performed analytically
and the result is
\be
\rlabel{CHPT-NF}
\Pids(q^2)_{CHPT-NF} =
\frac{B_0^2 F_0^2}{16\pi^2}\left[\frac{4\mu^2}{q^2}
+ 1 \right] \ .
\ee
The quartic dependence on $\mu$ cancels between the different diagrams as
required by chiral symmetry.
Therefore we get for the $B_K$ factor
\be
\rlabel{bknlo}
B_K(\mu)_{CHPT} =
\frac{3}{4}\left( 1 -\frac{1}{16\pi^2 F_0^2}
\left[2 \mu^2 + \frac{\dis q^2}{\dis 2}\right] \right)\ .
\ee
The correction is negative. It disagrees somewhat with the result obtained
in \rcite{BBG1}
because there no attempt at identifying the cut-off
across different diagrams was made.
Since we work at leading level in $1/N_c$
in the NLO CHPT corrections we have included the relevant
singlet ($\eta_1$) component as well using nonet symmetry.
The correction in \rref{bknlo} has precisely
the right behaviour to cancel partly $C(\mu)$ which increases with
increasing $\mu$. We will defer a numerical discussion till later.

\section{The NJL model}

The Lagrangian of the NJL model is given by
\begin{eqnarray}
\rlabel{LENJL}
{\cal L}_{NJL} &=&
\overline{q}\left[ i\gamma^\mu
\left(\partial_\mu - i v_\mu -i a_\mu \gamma_5 \right)
-{\cal M} - s + ip\gamma_5\right] q \nonumber\\
&&+ \, 2 g_S \sum_{i,j} (\overline{q}^i_R q^j_L )(\overline{q}^j_L q^i_R)
\end{eqnarray}
with $\overline{q} = (\overline{u}\ \overline{d}\ \overline{s})$,
$i$, $j$ are flavour indices and
${\cal M} = \mathrm{diag}(m_u , m_d , m_s)$. The coupling
$g_S$ is  given by
$g_{S} = $ $4\pi^2$ $G_{S}/(N_c \Lambda_\chi^2)$
in terms of the notation used in \rcite{BRZ,BBR,BP2}.
$v_\mu,a_\mu, s$ and $p$
are external vector, axial-vector, scalar and pseudoscalar fields. These
are used to probe the theory with.
This Lagrangian can be argued to follow from QCD in the following way:
all indications are that in the pure glue sector there is a mass gap. The
lowest glueball in lattice QCD numerical simulations
has  a mass of about 1.5 GeV. This means that correlations below
this scale should vanish. We then treat the interactions of quarks
below this scale as pointlike. It can also be seen as the first term
 in an expansion in local terms after integrating out fully the gluons.
The Lagrangian in \rref{LENJL} has the
correct chiral symmetry properties. Thus results calculated within this
formalism will have the correct chiral properties.

This model, for $G_S > 1$, spontaneously develops a quark vacuum expectation
value so that the chiral symmetry is spontaneously broken.
We will now work within the $1/N_c$ expansion in this model.
How to calculate two-point functions and three-point functions in the presence
of non-zero current masses can be found in \rcite{BP2}. We have for this
work evaluated all necessary two-, three- and four-point functions in terms of
different masses. The prescription used for calculating the integrals
is the same as in \rcite{BP2}. We first use identities
of the type $2 q\cdot r = ((q+r)^2 - M^2) - q^2 - (r^2 - M^2)$ and
$r^2 = (r^2 - M^2) + M^2$ to reduce all integrals to
scalar loop integrals. In these, we then combine the propagators
using Feynman parameters. Then we rotate to Euclidean space and
regularize the integrals using a consistent proper time cut-off
which introduces a cut-off scale $\Lambda_\chi$ in all the loops
with constituent quarks. This procedure reproduces the two-point
functions of \rcite{BRZ} and in a
low-energy expansion all results of \rcite{BBR}.
The flavour anomaly is also correctly
treated in this way, see \rcite{BP3}.

As input parameters we use those obtained in the best fit to $F_0$ and the
$L_i$ of CHPT in \rcite{BBR} for the full ENJL model, i.e. including
the vector--axial-vector four-quark term which is modulated by the
coupling $G_V$. At present we are mainly
interested in seeing the effect of SU(3) breaking and we shall
put $G_V$ to zero, i.e. we work within the model in eq.
\rref{LENJL}. The use of the same input values here as the case
with $G_V\ne 0$ will also allow to get a better comparison
with that more realistic case which is in preparation.
These input values are
$G_S = 1.216$ and $\Lambda_\chi = 1.16$ GeV.
This gives a constituent quark mass
of about 265 MeV in the chiral limit. We will use $m_d = 3.2$ MeV
and $m_s = 83$ MeV. These are the values that reproduce the
experimental pion and
kaon mass using the full ENJL model.

In the NJL model of eq. \rref{LENJL}, these input parameters correspond to
the quark condensate $\langle \overline{q} q \rangle_{\rm{Chiral}}
\approx - (281 \, \rm{MeV})^3$,
$F_0\approx 114$ MeV, $B_0\approx 1.7$ GeV and
$m_K^2 \approx 0.133$ GeV$^2$. We will also study the case where both
quark masses are equal because this is where the lattice calculations have
been done so far,
then we shall use $m_s = m_d = 43$ MeV. This leads to the same kaon mass
as given above.

\section{The NJL calculation}

We now proceed to evaluate $\Pids$ in the NJL model.
The leading contribution comes from the class of diagrams
in fig. \tref{fig2}.
\begin{figure}
\epsfxsize=14cm 
\epsfbox{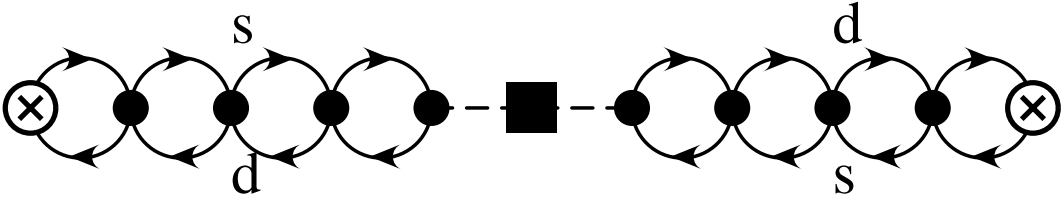}
\caption{The leading $1/N_c$ contribution to $\Pids(q^2)$
in the NJL model.
Symbols as in fig. \protect{\ref{fig1}} except that a dot is now a NJL-vertex
and the full lines are constituent quark-lines.
The flavour is mentioned next to
the lines.}
\rlabel{fig2}
\end{figure}
In terms of the mixed pseudoscalar-axial-vector two-point function
$\Pi_{P\mu}(q)_{sd}$ (see \rcite{BRZ} for definition), the result is
\be
\rlabel{NJLLO}
\Pids(q^2) = - \frac{\dis 1}{\dis 2}
\Pi_P^\mu(-q)_{sd} \Pi_{P\mu}(q)_{sd}\ .
\ee
This mixed two-point function  has been calculated in the full ENJL model
 in the chiral limit \rcite{BRZ} and in \rcite{BP2} for non-zero
quark masses to all orders in the chiral expansion. The lowest order
term in the expansion of the result
in \rref{NJLLO} coincides with that in \rref{CHPT-LO}
when the parameters are those determined from the NJL model.

At the non-factorizable level life becomes a lot more complicated.
There are two major classes of diagrams in the leading contribution
to the non-factorizable case. They are depicted in figs. \tref{fig3}
and \tref{fig4},
we will refer to these as 3-point and 4-point diagrams respectively.
\begin{figure}
\epsfxsize=14cm 
\epsfbox{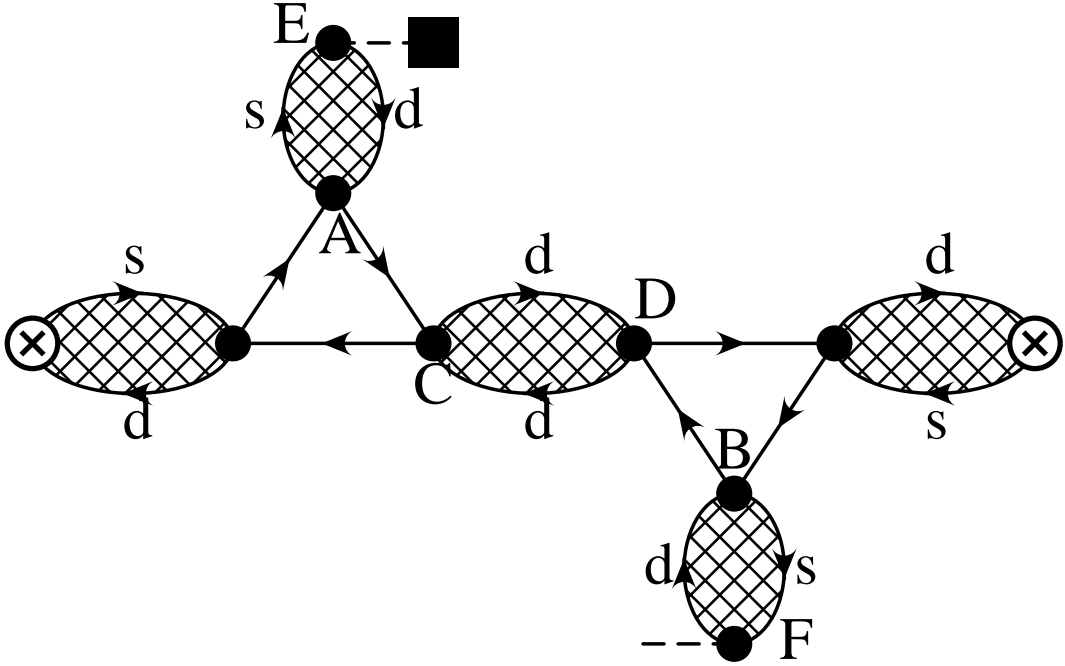}
\caption{The class of 3-point diagrams. Symbols as in fig.
\protect{\ref{fig2}},
the hatched areas are a summation over sets of one-loop diagrams
as shown in fig. \protect{\ref{fig2}}.}
\rlabel{fig3}
\end{figure}
\begin{figure}
\epsfxsize=14cm 
\epsfbox{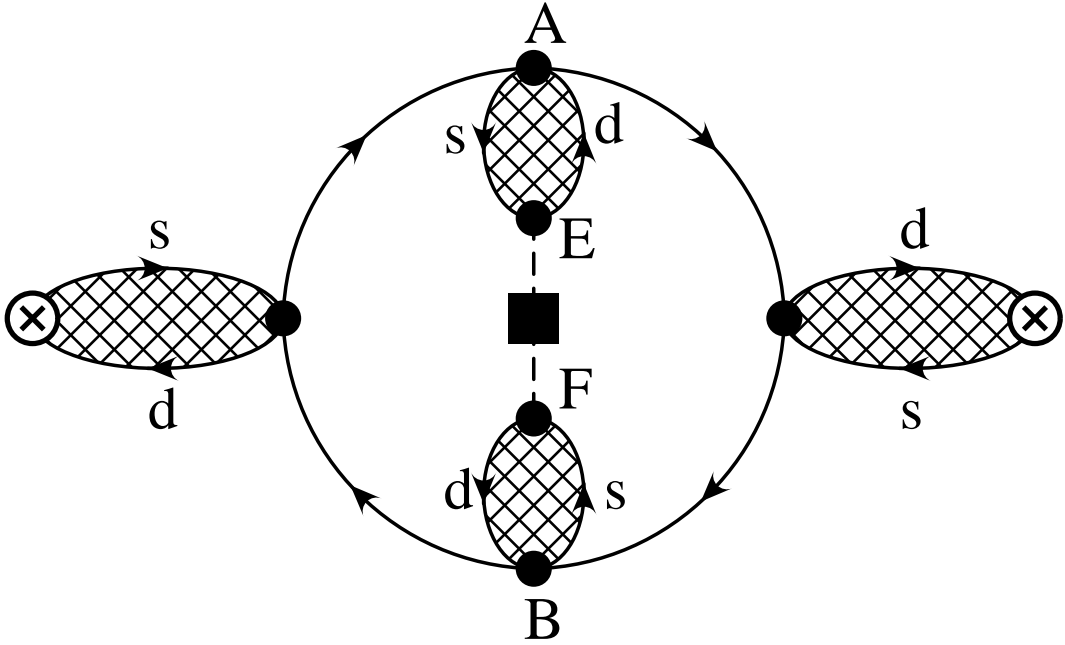}
\caption{The class of 4-point diagrams, symbols as in fig.
\protect{\ref{fig3}}.}
\rlabel{fig4}
\end{figure}
 In addition the diagram of fig. \tref{fig3} also exists
with everywhere the quarks s and d
interchanged and the directions of all fermion
lines reversed. The latter diagram can then related to the first version
by CPS-symmetry and can be calculated using the same code written
for the diagram of fig. \tref{fig3} but with the quark masses
$m_s$ and $m_d$ interchanged. We have checked that diagram
in fig. \tref{fig4} gives the same result after that interchange.
The tails here can be 0,1,2,3,... loops and they can be resummed using
the procedure given in \rcite{BRZ}. The expressions needed for the
tails are given in \rcite{BP2}. They represent in a sense the fermionic
equivalent of the propagators of the mesons. Notice that the two-point
functions needed also contain parts that do not have a pole.
Here it can be seen where the main complexity of the calculation comes from.
The $X$ boson
has both vector and axial-vector couplings, the internal coupling
of the tail to the 3 or 4 point quark loop can thus have 4 different Dirac
structures. The three- and four-point
functions themselves have a rather complicated
momentum dependence and need an explicit numerical integration over the
Feynman parameters to be evaluated. This makes it rather important
to have several consistency checks on the final numerical result. We will
describe the ones we have used below in addition to the interchange of
masses on the 4-point diagram.
We refer to the 3-point diagrams by the form of the insertion at
places A,B,C and D and the 4-point ones by their insertions at A and B.
This in turn fixes the insertion in the points E and F in both diagrams.
As an example let us quote the expression for the PSSP 3-point
diagram corresponding to the flavours in fig. \tref{fig3},
\ba
\mathrm{<PSSP>} &\equiv&
\frac{\dis i}{\dis 2} \left[1+g_S \Pi_P(q)_{sd}\right]^2 \,
g_S^3 \, \int^\mu_0 \frac{d^4 r}{(2\pi)^4} \,
\Pi_P^\mu(-q-r)_{ds} \Pi_{P\mu}(q+r)_{ds} \nonumber \\
&\times &  \overline{\Pi}^{sdd}_{PPS}(q+r,-r)
\left[1+g_S\Pi_S(-r)_{dd}\right] \,
\overline{\Pi}^{sdd}_{PPS}(-q-r,r)\ .
\ea
The others are considerably more involved. The two- and
three-point functions here are defined with the notation used in \rcite{BP2}.
The non-barred ones correspond to the full functions and the
barred ones to the one loop expressions\rcite{BRZ,BP2}.
We have made sure that the expressions for the underlying 3 and 4-point
one-quark-loop expressions obey all the Ward identities that follow
from the symmetry in the one-loop case
for the most general case, all masses different. The current identities
are satisfied with the constituent masses. At the present level there
is no contribution from diagrams that produce a Levi-Civita symbol. These
only contribute at the level considered here as soon as $G_V\ne 0$.
A large fraction of the work was involved in making sure the 3- and 4-point
one-loop functions were correct.

\section{What happened to the Fierzed term ?}

In the standard VIA we have
$B_K(\mu)=\left(3/4\right)\left(1+ 1/N_c\right)$.
{}From the diagram in fig. \tref{fig2} we get
$B_K (\mu) = 3/4$ and the difference comes from the colour suppressed
combination of attaching
the operator $\Gamma_{\Delta S=2}$ to the quark lines.
In terms of the $X$ boson propagator this type of contributions corresponds
to taking the diagram of fig. \tref{fig4} and replacing the lines connecting
A-E and B-F by a pointlike insertion and sending the cut-off $\mu$ to
infinity. We can then rearrange the integration over $r$ and
the internal quark loop momenta such that the new integration variables
have no mixed dependence in any of the quark propagators.
We can now split the double integral into two independent
parts. The expression for this diagram then reduces to the one
in fig. \tref{fig2} with an extra factor of $1/N_c$ in front
obtaining the usual VIA result.
One can wonder how good is the VIA (i.e., the Fierzing)
when one introduces the
cut-off $\mu$ in the $r$ momentum. The answer is that this $1/N_c$
is only reached for rather high values of the cut-off $\mu$.

Let us also see what happens in the context of chiral perturbation theory.
In fact the same contribution appears in the CHPT
calculation. It is the one depicted in fig. \tref{fig1}d. To the order
we have calculated here (i.e. next-to-leading order)
it is exactly zero, the vector-vector and
axial-vector--axial-vector cancel exactly. It is only
at higher orders in the CHPT expansion, when the vertex
vector(vector-axial)--$X$ boson is  the one coming from the
$L_{10}$ term in the ${\cal O}(p^4)$ chiral Lagrangian,
when this contribution starts
showing itself. This in fact allows us a rare check on a single
diagram. The chiral calculation for the
vector(axial-vector)--vector(axial-vector) piece gives
(at NLO) a contribution of
\be
{\Pids}^{Fierz}_{VV(AA)} (q^2) =
(-) \, \frac{B_0^2 F_0^2}{16\pi^2} \, 4 \, \frac{\mu^4}{q^4}\ .
\ee
In the NJL calculation,
this agrees for small values of $\mu$ (up to about 0.35 GeV)
well. For larger values of $\mu$ the contribution of
this diagram
is $(3/4) \, \epsilon$, with
$\epsilon \approx$ 0.006, 0.03, 0.08, 0.1, 0.2 and 0.3
 for $\mu= $ 0.3, 0.5, 0.7, 0.9, 1.1 and 1.5 GeV.
 As can be seen the Fierz term is slightly
suppressed for the values of $\mu$ that are relevant in the next section.

\section{Numerical results}

We have studied three cases, namely, the chiral case mentioned above;
the case with SU(3) symmetry breaking $m_s\ne m_d$  with the input
parameters introduced in section 3 and the case with non-zero quark
masses but with $m_s=m_d$ also introduced in section 3.
The procedure we have followed to analyze the numerical results
is the following.  We fit the ratio between the correction and the
leading $1/N_c$
result for a fixed scale $\mu$ to $a/q^2 + b + c q^2$ which always gives
a very good fit ($a$, $b$ and $c$ are $\mu$ dependent).
 Once we have this fit we can extrapolate our
$B_K$ form factor (remember that we have calculated it for
Euclidean $q^2$) to the physical $B_K$, i.e. to $q^2=m_K^2$.

Let us first treat the chiral or massless quarks case.
Here a nontrivial check on the
results is that the diagrams have a behaviour which sums to $1/q^2$.
The individual contributions do not have this behaviour. In fact for
our lowest $q^2$ the cancellations are rather significant.
This cancelation happens point by point in the loop momentum
($r$) integral since there is an
underlying chiral relation for the 4-point amplitude.
Therefore we have used a deterministic numerical integration scheme
to have all diagrams evaluated at the same internal momenta. This increases
the accuracy in the sum beyond the accuracy in the separate diagrams since
only the noncanceling contributions survive at all stages of the calculation.
We also see here a small $q^2$ dependence in the $B_K$ form
factor, this is
for $B_K$ as defined above but with off-shell kaons.

In this case $a$ is compatible with zero as required by chiral symmetry. $b$
is the relevant contribution to $B_K$ since $m_K^2|_\chi=0$.
The first three
columns in table \tref{table1} are $\mu$, $B_K^\chi(\mu)$ and $\hat B_K^\chi$
for this case. For calculating $\hat B_K$ we use
$a_+ = -2/9$ and $\Lambda^{(3)}_{\overline{MS}} = 250$ MeV
which corresponds to $\alpha_S^{(1)}(1.5 \ {\rm GeV})=0.39$.
\begin{table}
\begin{center}
\begin{tabular}{c|cc|ccc|cc}
$\mu$ (GeV) & $B_K^\chi(\mu)$ & $\hat B{}_K^\chi$ & $B_K^m(\mu)$ &
$B_K^a(\mu)$ &
 $\hat B_K^m$ & $B_K^{\rm eq}(\mu)$ & $\hat B_K^{\rm eq}$ \\
\hline
0.3  & 0.68 &0.50 & 0.74&   0.50&0.55  &0.74 &0.55 \\
0.5  & 0.59 &0.59 & 0.71&$-$0.44&0.71  &0.72 &0.72 \\
0.7  & 0.53 &0.58 & 0.69&$-$2   &0.75  &0.68 &0.75 \\
0.9  & 0.48 &0.55 & 0.66&$-$3   &0.76  &0.65 &0.75 \\
1.1  & 0.45 &0.54 & 0.64&$-$4   &0.76  &0.64 &0.76
\end{tabular}
\end{center}
\caption{Results for $B_K$ and $\hat B_K$.}
\rlabel{table1}
\end{table}
For very small $\mu$ we also see agreement with the CHPT calculation
given above with the values for $B_0$ and $F_0$ expected for our NJL input
values. The matching with QCD leading logarithmic evolution, i.e.
the stability of $\hat B_K$, is not very good in this case.

In the second case ($m_s \ne m_d$),
because the chiral symmetry is broken, there is
a possibility for contributions to $B_K(q^2)$ that are not
proportional to $q^2$, i.e. $a\ne 0$.
In fact a CHPT calculation like the one above predicts
precisely the presence of this type of terms. For
small values of $q^2$ the part due to $a$ dominates even though it is only a
small correction when extrapolating to the physical $B_K^m$ at $q^2=m_K^2$.
This can be found in column 4.
The fifth column is the form factor $B_K^a$ for $q^2=-0.001$
 GeV$^2$ where the correction due to the $a$ term is sizeable.
Notice the difference
between these two columns.
This same feature should be visible in the lattice
calculations as soon as they are done with different
quark masses. The invariant $\hat B_K^m$ for this case is in column 6.

In the the last case, i.e. $m_d=m_s$, which is similar to the present
lattice QCD calculations,  the fit gives $a=0$ to a good
precision and  the value of $B_K^{\rm eq}$ extracted is rather
independent of $q^2$. The invariant $\hat B_K^{\rm eq}$ in this case is in
column 8.

A reasonably stable value for $\hat B_K^\chi$
occurs for scales $\mu\approx (0.7 \sim 1.1)$ GeV.
But the difference with the massless case is numerically significant.
We obtain
\be
\hat B_K^m(m_K^2\approx 0.13~{\rm GeV}^2) \approx 1.35
 \, \hat B_K^\chi(m_K^2= 0)
\ee
for scales $\mu\approx (0.7\sim 1.1) ~{\rm GeV}$.
\section{Conclusions}
We have studied a nonleptonic weak matrix element in a way where the short
and long distance contributions are separated in a physically visible
fashion through the $X$-boson propagator.
This allows us to identify the r\^ole
of $\mu$ across diagrams and to connect with the running in the short distance
domain. We have done this in the NJL-model to all orders in momenta there.
This gave us a result that has manifestly the correct chiral properties.

In view of the results of \rcite{BBR,NJL,BP2} we expect to get
a good prediction for the effects of non-zero and different current quark
masses.
We see those and find a significant
change due to both. For the extrapolation to the kaon pole the
difference between the masses has a much smaller effect than the fact that
they were non-zero. In order to compute $B_K$ in the general case a careful
extrapolation to the poles was needed. The final correction to the $B_K$
parameter compared to its leading value of $3/4$ turns out to be rather
small. The stability of $\hat B_K$ when quark masses are
present, is around scales $\mu \approx (0.7 \sim 1.1)$ GeV.
For these scales one may expect that the contribution of
spin 1 mesons (remember that only the spin 0 mesons are the
ones included for $G_V=0$) is important. This may be implemented
like in the case of
the $\pi^+ -\pi^0$ mass difference \rcite{BBG2,BR,BRZ} by going
to the full ENJL model with vector--axial-vector mesons included.
This also will allow us to compare with the results of the
next-to-leading $1/N_c$ corrections obtained in similar
approaches \rcite{Gerard}.
We are studying now this case \rcite{BP1}.
We are also studying whether we can use the present program to
obtain more information about higher order terms in the chiral Lagrangian.

\section*{Acknowledgements}
We thank Eduardo de Rafael for discussions.
This work was partially supported by NorFA grant 93.15.078/00.
JP thanks the
Leon Rosenfeld foundation (K\o{}benhavns Universitet) for support and
CICYT(Spain) for partial support under Grant Nr. AEN93-0234.

\end{document}